\documentclass[preprint,12pt]{elsarticle}
\usepackage{amsmath}
\usepackage{amssymb}
\usepackage[section]{placeins}
\usepackage{units}
\usepackage{graphicx}

\newcommand{\pt}{p_\perp}

\newcommand{\taui}{\tau_\text{i}}

\newcommand{\jewel}{\textsc{Jewel}\ }

\journal{Physics Letters B}

\begin{document}


\begin{frontmatter}

\title{Geometrical aspects of jet quenching in JEWEL}

\author{Korinna C. Zapp}

\address{Department of Physics, CERN, Theory Unit, CH-1211 Geneva 23}

\begin{abstract}
In this publication the performance of the Monte Carlo event generator \textsc{Jewel} in non-central heavy-ion collisions is investigated. 
\textsc{Jewel} is a consistent perturbative framework for jet evolution in the presence of a dense medium. It yields a satisfactory description of a variety of jet observables in central collisions at the LHC, although so far with a simplistic model of the medium. Here, it is demonstrated that also jet measurements in non-central collisions, and in particular the dependence of the jet suppression on the angle relative to the reaction plane, are reproduced by the same model.
\end{abstract}


\end{frontmatter}

\section{Introduction}

During the first years of the LHC operation the experiments ALICE, ATLAS and CMS have analysed the properties of jets emerging from heavy-ion collisions in great detail~\cite{Abelev:2013kqa,:2012ch,Reed:2013rpa,Aad:2010bu,Aad:2012vca,ATLAS-CONF-2012-115,Aad:2013sla,Chatrchyan:2011sx,Chatrchyan:2012nia,Chatrchyan:2012gw,Chatrchyan:2013kwa}. This wealth of data is a challenge to our understanding of jet quenching including as yet unsolved questions, for instance concerning the back-reaction of jets on the medium. Understanding how jets evolve and interact in a medium may ultimately give insights into the transition between weakly and strongly coupled regimes and reveal properties of the medium not accessible to other probes.

At LHC energies hard jets are produced copiously and a considerable fraction of the fragments is accessible above the soft background facilitating detailed studies of the jet structure, fragmentation functions etc. In this situation the developments of Monte Carlo event generators for jet quenching both as a theoretical and an experimental tool is essential. On the theory side it is currently the only technique allowing for the calculation of exclusive final states thus giving access to the structure of jets. They also facilitate a detailed comparison to experimental data. On the experimental side simulation tools are needed to correct for acceptance, efficiency etc. and to determine the transfer matrices needed for unfolding of detector effects.

\jewel~\cite{Zapp:2012ak,Zapp:2013vla} is a publicly available Monte Carlo event generator for jets in heavy ion collisions. It is based consistently on perturbative language to describe jet evolution and interactions in a medium in a common framework. By construction limitations to analytic approaches such as kinematic limitations, momentum conservation, restriction to single gluon emission etc. are overcome. So far \jewel has been shown to reproduce a number of very different jet quenching data for central collisions rather satisfactorily. In this publication the centrality and azimuthal dependence of these observables is studied.

\section{Jet quenching in JEWEL}

In this section a short summary of the physics of \jewel is given, a detailed discussion can be found in~\cite{Zapp:2012ak}.

Hard scatterings of composite objects such as protons resolve the partonic structure of the interacting objects even if the objects at their own characteristic scale cannot be described in a perturbative language. According to (proven and postulated) factorisation theorems the non-perturbative structure has no influence on the hard interaction. We apply the same reasoning to hard interactions of a jet in a quark-gluon plasma. This implies that such hard interactions can be described with standard perturbative techniques. 

The assumptions underlying the \jewel construction are that (i) the interactions of jets resolve quasi-free partons in the medium, (ii) an infra-red continued version of the perturbative scattering matrix elements can be used to describe all interactions of jets in the medium, (iii) the formation times govern the interplay of different sources of radiation and (iv) the LPM effect can be included by generalising the probabilistic formulation in the eikonal limit to general kinematics.

Thus, in \jewel leading-order matrix elements and parton showers are used not only for the initial production of hard jets, but also for the re-scattering of jets off partonic constituents of the medium. In the case of hard re-scattering with a mean free path longer than the time needed for the parton shower evolution this approach certainly makes sense. The extension of this picture into the regime of semi-hard and soft re-scattering is an assumption (corresponding to assumptions (i) and (ii)). The benefit of combining in this way LO matrix elements with parton showers is that both elastic ($2\to 2$) and inelastic ($2\to n, n>2$) processes are generated with the leading-log correct relative rates\footnote{This approach can be promoted to higher accuracy by combining several real-emission matrix elements~\cite{Catani:2001cc,Mangano:2006rw}, using NLO matrix elements~\cite{Frixione:2002ik,Nason:2004rx,Hoeche:2012yf} or including more next-to-leading log terms in the parton shower~\cite{Gustafson:1987rq}. However, as the dominant uncertainty comes from the infra-red continuation, it is not necessary in this case.}. 

When re-scattering and radiation take place on comparable time scales it is possible that several sufficiently hard scattering processes (including the initial hard jet production) can induce radiation simultaneously. Assumption (iii) states that in this case only the emission with the shortest formation time can be formed while the others are discarded. Since the formation times are correlated with the hardness of the emission, this statement can be rephrased in terms of momentum scales: on average, the hardest emission will be formed. As the scale of the first emission is determined by the scale of the $2\to 2$ scattering process, this means that only re-scatterings that are harder than the virtuality of the hard parton can emit radiation. It is therefore very unlikely that a re-scattering in the medium, which tends to be soft or semi-hard, can disturb the evolution of a highly virtual parton. This was also found independently in calculations of medium induced radiation off colour dipoles and discussed in~\cite{CasalderreySolana:2012ef}. There, the authors also consider coherent radiation off an ensemble of unresolved partons, which can in principle be included in \jewel as well but is not part of the current implementation. The advantage of the Monte Carlo formulation is that the interplay of different processes is generated fully dynamically.

When several scattering processes take place within the formation time of an induced emission they are known to act coherently (LPM-effect). This quantum mechanical interference can be effectively taken into account in a probabilistic framework by a self-consistent determination of the number of contributing processes and the kinematics of the emission (which determines the formation time). In addition, the emission probability has to be adjusted~\cite{Zapp:2008af,Zapp:2011ya}. This prescription has been generalised from eikonal to general kinematics and is included in \textsc{Jewel} (assumption (iv)).

\smallskip

Thus, in Jewel all radiation is generated by parton showers and it is
in general not possible to ascribe an emission to the evolution associated
with the original hard jet production or a re-scattering. The splitting kernels are not modified by the presence of the medium, but the parton shower radiates more than in vacuum because sufficiently hard re-scattering effectively restarts the QCD evolution at a higher scale.

\medskip

The initial jet production matrix elements and initial state parton showers are simulated with \textsc{Pythia}\,6.4~\cite{Sjostrand:2006za} using the EPS09 nuclear PDF set~\cite{Eskola:2009uj} on top of the  \textsc{Cteq6l1}~\cite{Pumplin:2002vw} set provided through the LHAPDF~\cite{Whalley:2005nh} interface. The final state parton shower evolution and re-scattering are simulated within \textsc{Jewel}. Finally, the events are handed back to \textsc{Pythia} for hadronisation and hadron decays. 

\section{A simple model of the medium}

For exploring and understanding the features and capabilities of this new approach to jet quenching it is advantageous to work with a simple model of the medium, so that one understands which features in the data can be accounted for by microscopic dynamics.  

A Glauber model~\cite{Eskola:1988yh} is used to relate centrality to impact parameter $b$ and to compute the density of binary nucleon-nucleon collisions $n_\text{coll}(b;x,y)$ and number of participants $n_\text{part}(b;x,y)$ in the transverse plane ($x$ and $y$ are the transverse coordinates, $z$ is the beam direction). Initial di-jet production is assumed to take place at $t=z=0$, the distribution in the transverse plane is given by $n_\text{coll}(b;x,y)$. The initial condition for the hydrodynamic evolution is determined by two parameters, namely the initial temperature $T_\text{i}$ in the centre ($x=y=0$) of a central collision ($b=0$) and the proper time $\tau_\text{i}$ at which the evolution starts. The transverse profile is fixed by assuming that the initial energy density $\epsilon(b;x,y,\taui)$ is proportional to the density of participants,
\begin{equation}
\label{Eq::epsi}
\epsilon(b;x,y,\taui) = \epsilon_\text{i} 
  \frac{n_\text{part}(b;x,y)}{\langle n_\text{part} \rangle (b=0)}
  \qquad \text{with} \qquad 
  \langle n_\text{part} \rangle (b=0) \approx \frac{2A}{\pi R_A} \,,
\end{equation}
where for simplicity a symmetric $A+A$ collision was assumed. Here, $R_A$ is the radius of the nucleus and $\epsilon_\text{i}\propto T_\text{i}^4$ is related to the initial temperature. This choice is motivated by the argument that soft particle production should scale with the number of participants while hard processes scale with the number of binary nucleon-nucleon collisions. The entire centrality dependence is defined through equation~(\ref{Eq::epsi}). The hydrodynamic evolution assumes Bjorken expansion~\cite{Bjorken:1982qr} neglecting transverse expansion and an ideal gas equation of state such that
\begin{align}
\epsilon(b;x,y,\tau) & = \epsilon(b;x,y,\taui)
         \left(\frac{\tau}{\taui}\right)^{-4/3} \,, \\
T(b;x,y,\tau) & \propto \epsilon^{1/4}(b;x,y,\taui) 
  \left(\frac{\tau}{\taui}\right)^{-1/3} \,.
\end{align}

For proper times earlier than the initial time $\tau_\text{i}$ of the hydrodynamic evolution the temperature is assumed to increase linearly with $\tau$. At very early times the jet evolution is still characterised by the high scales set by the initial jet production such that it is protected from disturbances due to re-scattering in the medium. There is thus very little sensitivity in \jewel to the assumptions about the very early phase of the medium.

\jewel only considers interactions in the deconfined phase, therefore  re-scattering is only possible as long as the local temperature is higher than the critical temperature $T_\text{c}$. When a re-scattering occurs the thermal parton is generated with flavour and momentum given by the thermal distribution of an ideal gas with the local temperature at the time and location of the scattering assuming vanishing chemical potential. An improvement compared to~\cite{Zapp:2012ak} is that the longitudinal boost for partons at $z\ne 0$ is taken into account in the momentum distribution. 

For the initial time and temperature the values found in~\cite{Shen:2012vn}  are chosen, namely $\tau_\text{i}=\unit[0.6]{fm}$ and $T_\text{i}=\unit[485]{MeV}$. The critical temperature is taken as $T_\text{c}=\unit[170]{MeV}$. 

While this simple model captures important characteristics (such as the rapid longitudinal expansion) of heavy ion collisions it misses certain other aspects, most importantly the transverse expansion. The aim of this work is to investigate to what extent \jewel is able to describe jet quenching observables differential in centrality with the simple medium model.

\section{Centrality and angular dependence of jet quenching}

\begin{figure}
\centering
\includegraphics[width=0.45\linewidth]{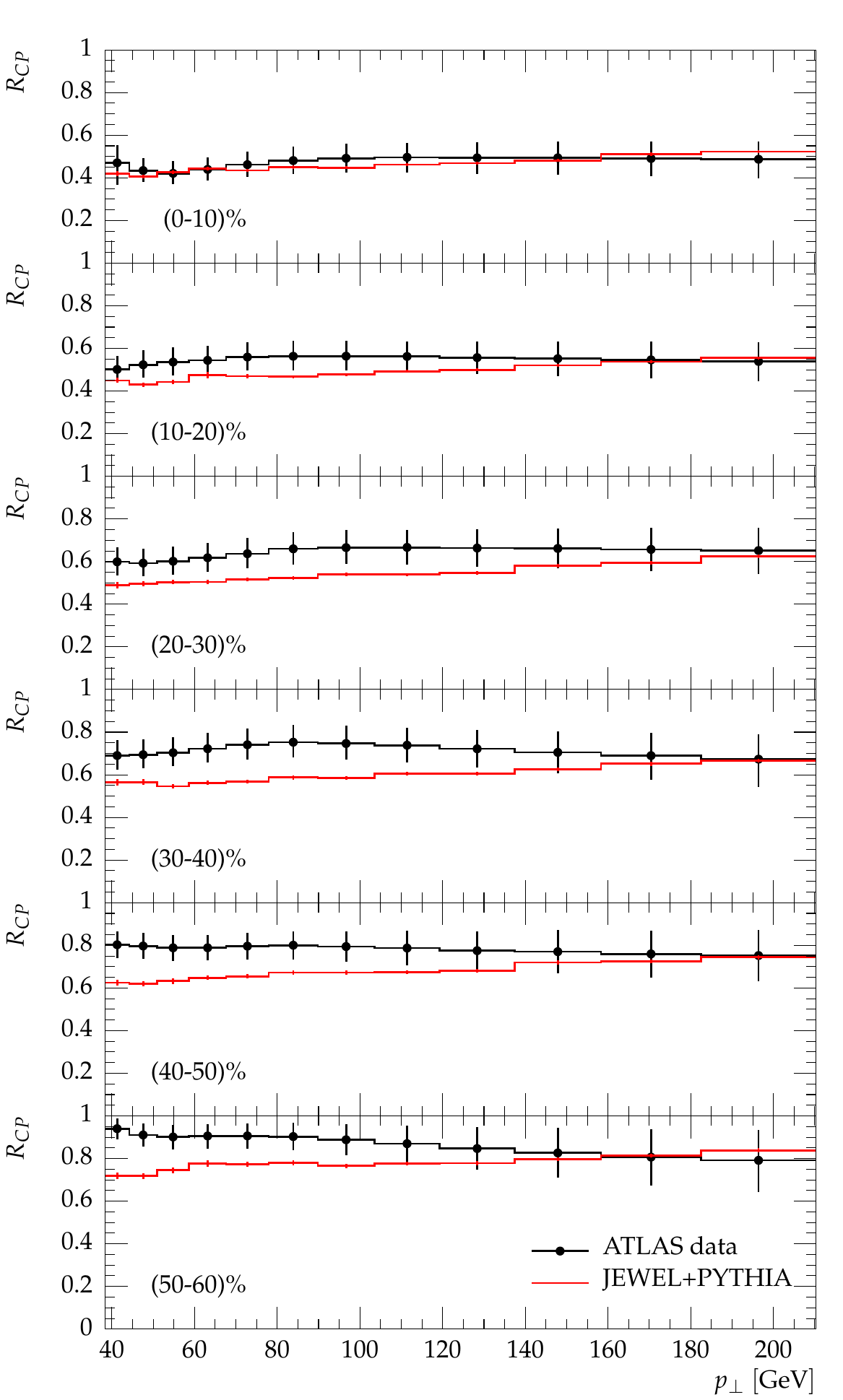}
\includegraphics[width=0.45\linewidth]{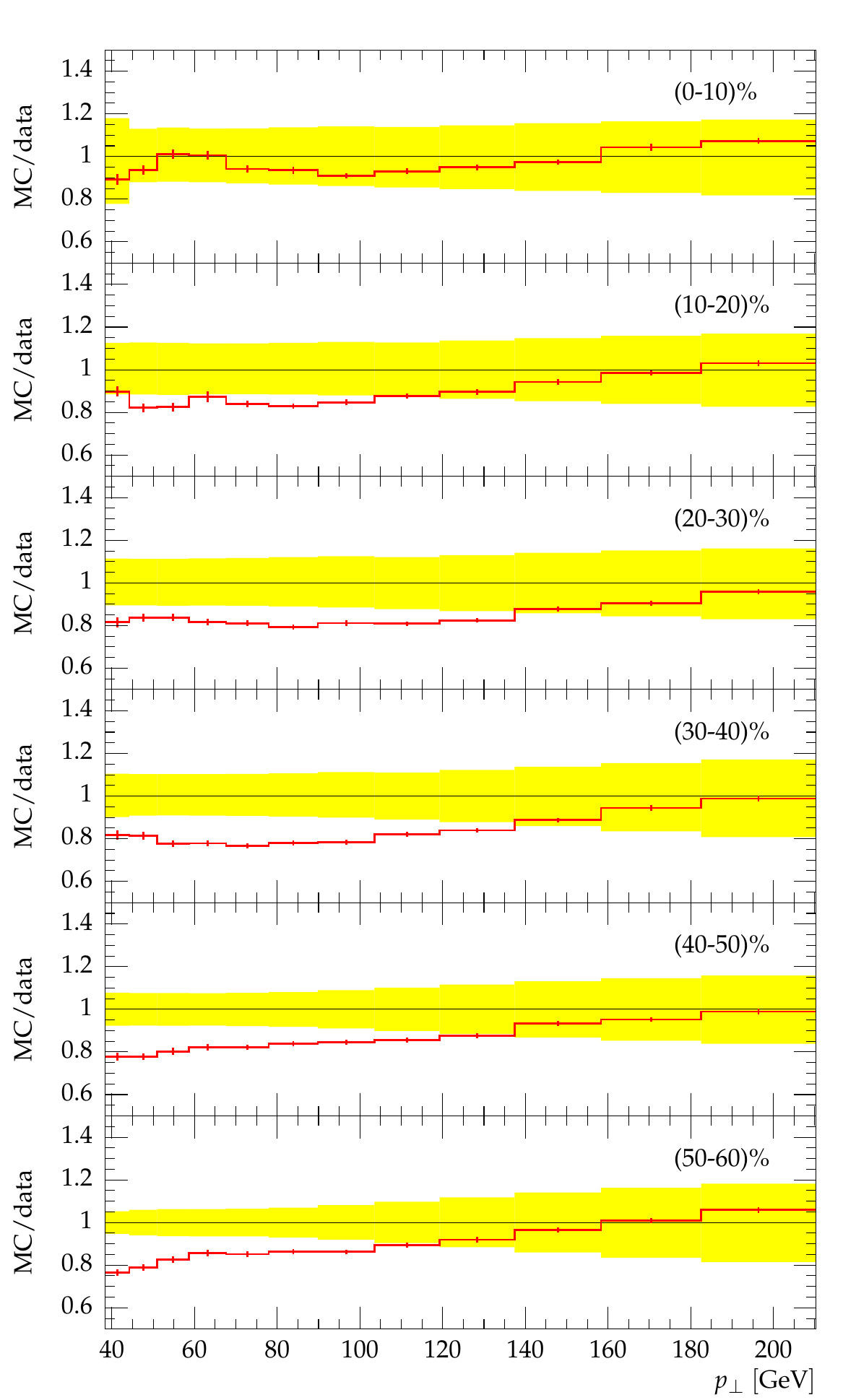}\\
\caption{Centrality dependence of the jet nuclear modification factor $R_\text{CP}$ in Pb+Pb collisions at $\sqrt{s_\text{NN}} = \unit[2.76]{TeV}$ for a jet radius $R=0.2$ and $|\eta_\text{jet}| < 2.1$~\cite{Aad:2012vca}. The $R_\text{cp}$ ratios are taken with respect to the (60-80)\% centrality class.}
\label{Fig::jetrcp}
\end{figure}

The results shown in this section were obtained with exactly the same settings as in~\cite{Zapp:2013vla}. There is only some freedom in choosing the exact value of the infra-red regulator $\mu_\text{D}$, in~\cite{Zapp:2013vla} it was found that a value of $\mu_\text{D}=2.7 T$ yields a reasonable description of a large variety of jet data in central collisions. Allowing for non-central collisions then does not introduce additional freedom. The analysis of Monte Carlo events was done with Rivet~\cite{Buckley:2010ar} using FastJet~\cite{Cacciari:2011ma}.

The centrality dependence of the inclusive jet suppression is shown in figure~\ref{Fig::jetrcp} for a small jet radius of $R=0.2$. The agreement between \jewel and the ATLAS data is very good in the most central bin and slightly worse in all others with a similar shape. The findings are similar for larger jet radii, although here the agreement is worse for central collisions (cf.~\cite{Zapp:2012ak,Zapp:2013vla}) and tends to grow better towards peripheral collisions (results not shown). Due to ambiguities in the treatment of background in data and the Monte Carlo the results for smaller jet parameters are presumably more reliable.

\begin{figure}
\centering
\includegraphics[width=0.45\linewidth]{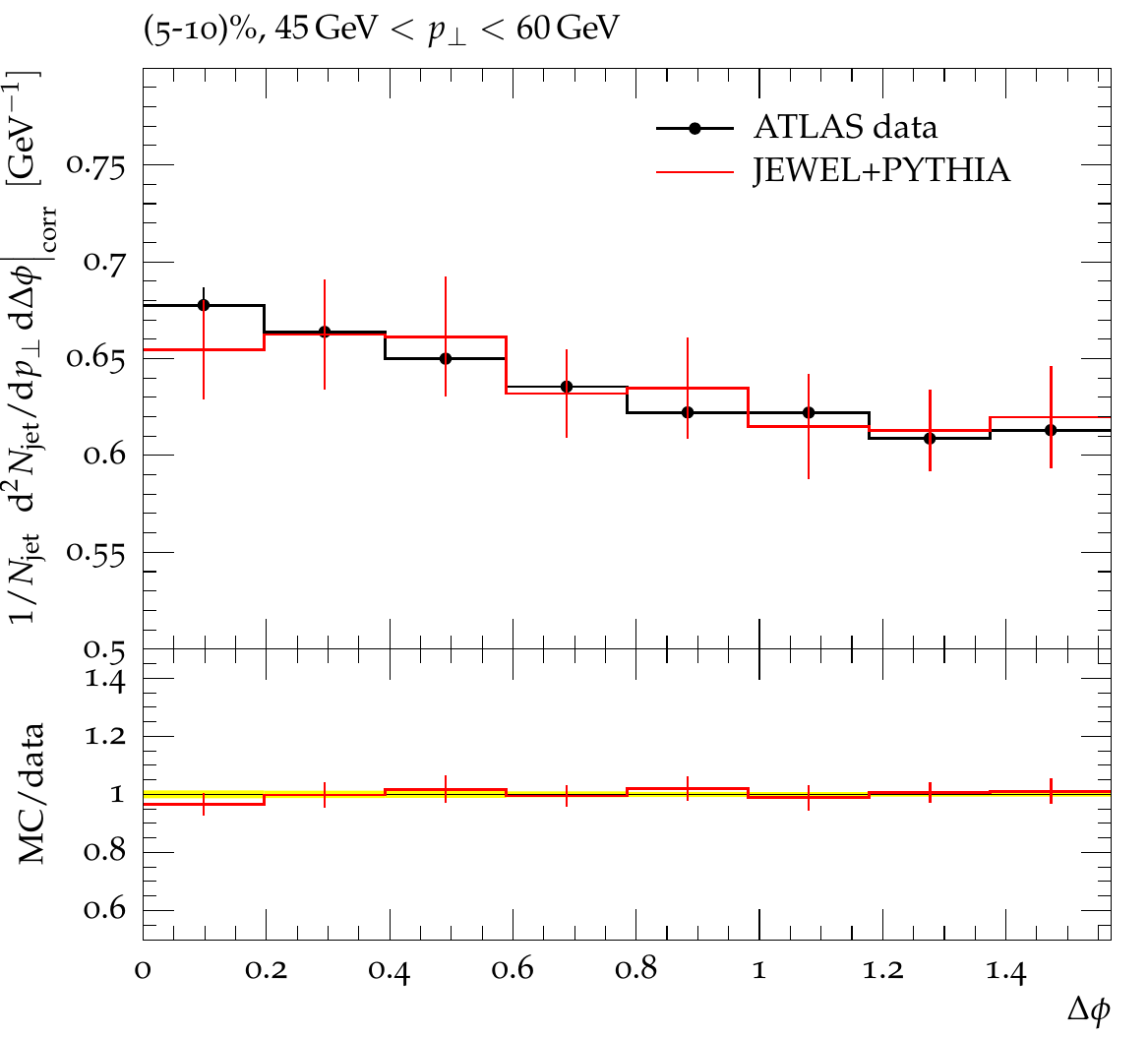}
\includegraphics[width=0.45\linewidth]{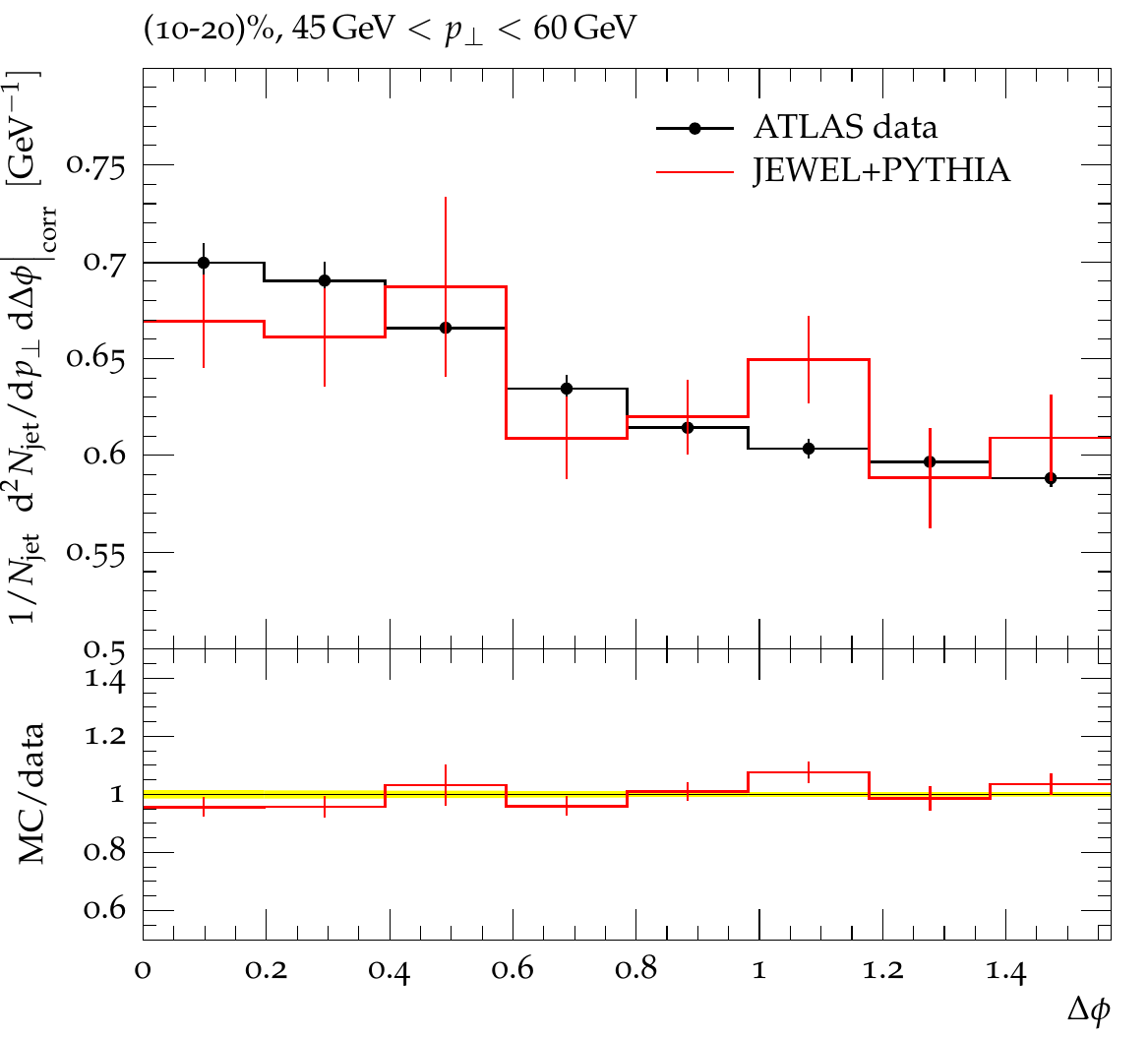}\\
\includegraphics[width=0.45\linewidth]{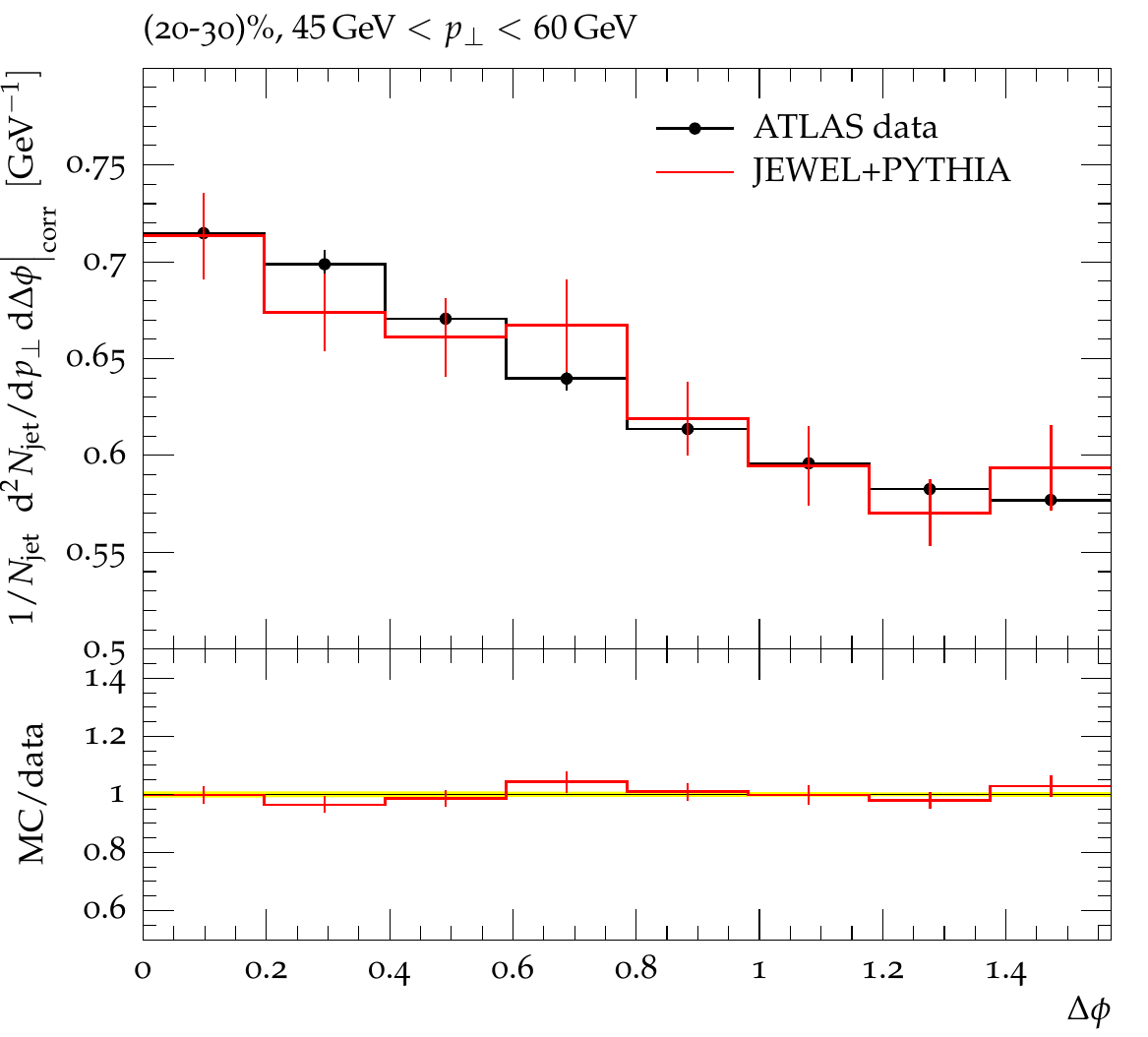}
\includegraphics[width=0.45\linewidth]{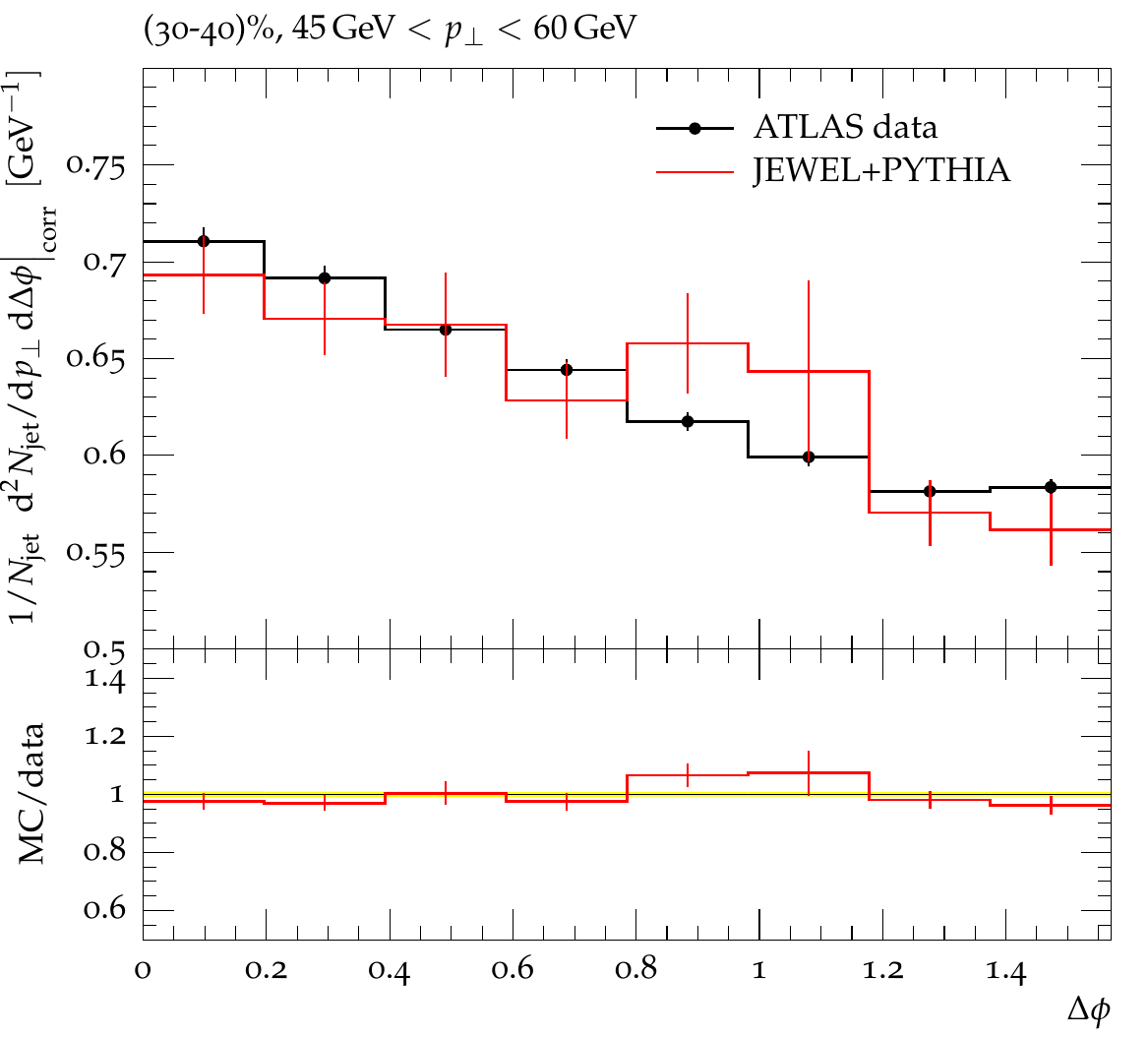}\\
\includegraphics[width=0.45\linewidth]{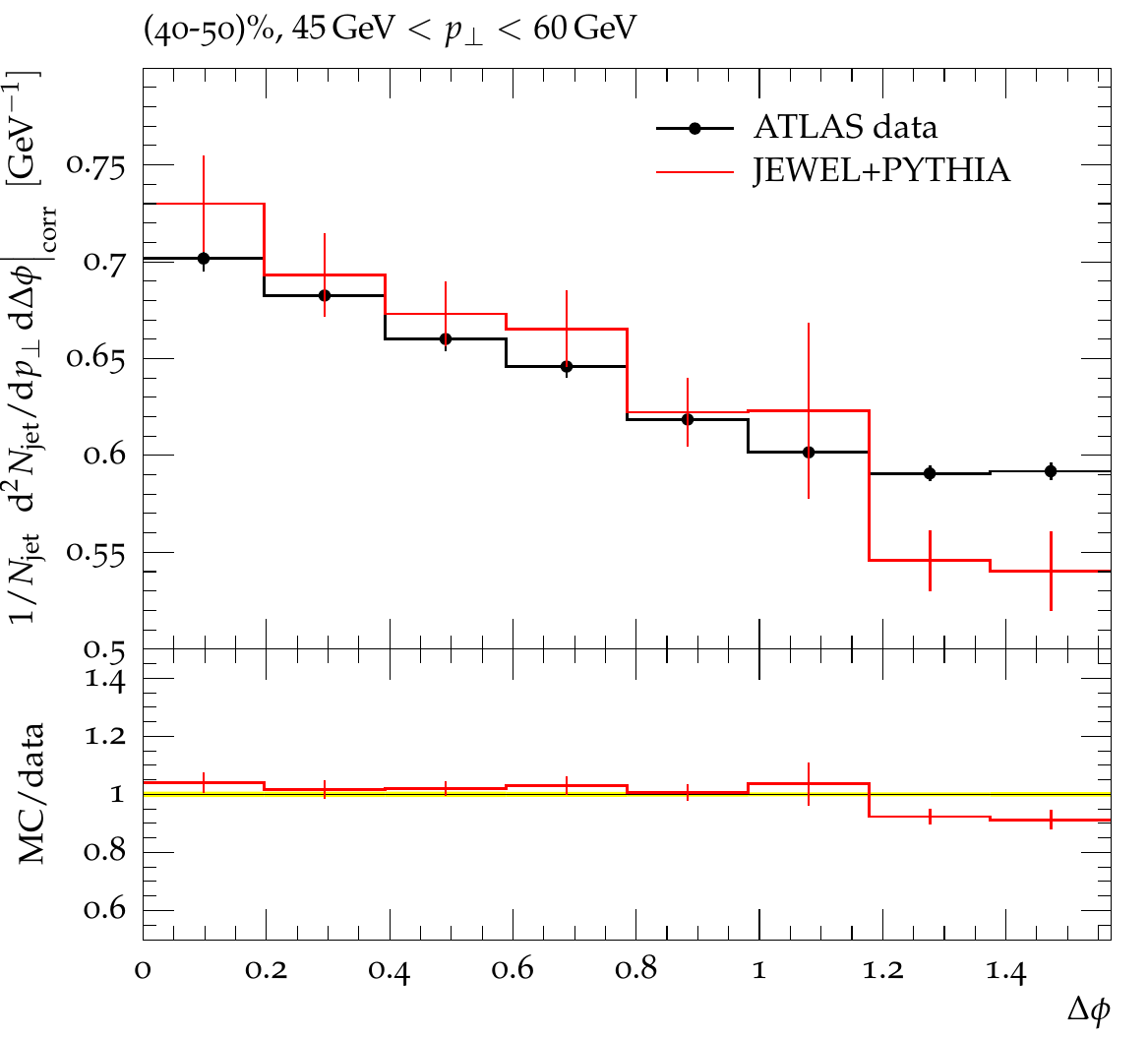}
\includegraphics[width=0.45\linewidth]{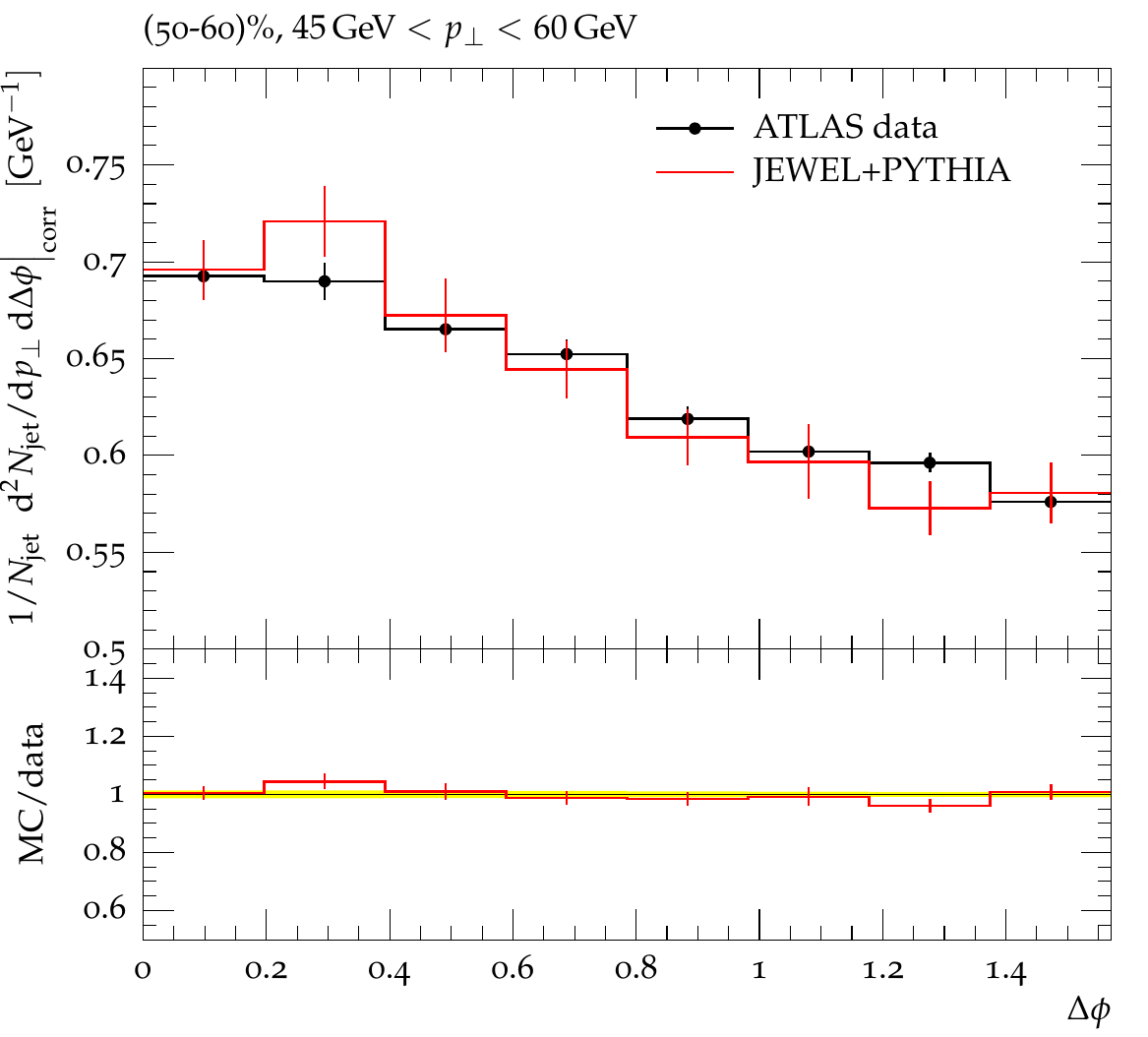}
\caption{Centrality dependence of the angular distribution of single inclusive jets in Pb+Pb collisions at $\sqrt{s_\text{NN}} = \unit[2.76]{TeV}$ for a jet radius $R=0.2$ and $|\eta_\text{jet}| < 2.1$ in the range $\unit[45]{GeV} < \pt < \unit[60]{GeV}$~\cite{Aad:2013sla} (data points read off the plots, only maximum of statistical an uncorrelated systematic errors shown). The data points have been corrected for event plane resolution using equation~(4) in~\cite{Aad:2013sla}.}
\label{Fig::phi}
\end{figure}

In figure~\ref{Fig::phi} the dependence of the single inclusive jet yield on the azimuthal angle relative to the reaction plane is presented across the entire centrality range. \textsc{Jewel+Pythia} describes these data very well, which is somewhat surprising given that the angular distribution is expected to be sensitive to the transverse expansion of the medium. Further studies will be needed to clarify this point. Figure~\ref{Fig::phi} only shows one bin in jet $\pt$, but the agreement between data and Monte Carlo is equally good in the other $\pt$ bins. This observable is complementary to the centrality dependence of the jet nuclear modification factor shown in figure~\ref{Fig::jetrcp}. The latter measures the overall suppression of jets in a given centrality class while the angular variation is only sensitive to the asymmetry of the medium (the distributions are normalised to the number of jets in the respective centrality class). While it is the same mechanism of jet quenching that is responsible for both effects, they are thus sensitive to different properties of the medium.

\begin{figure}
\centering
\includegraphics[width=0.45\linewidth]{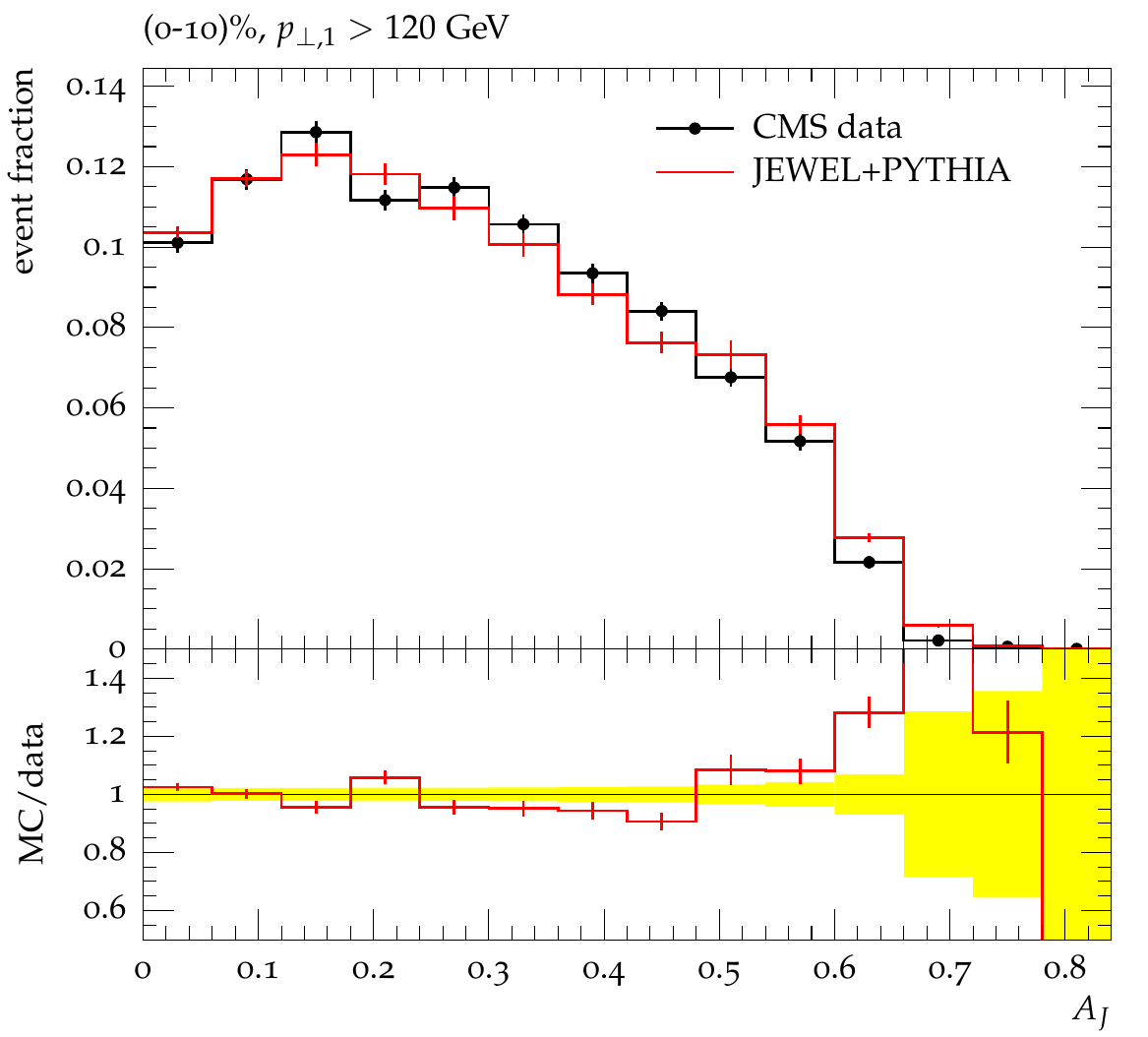}
\includegraphics[width=0.45\linewidth]{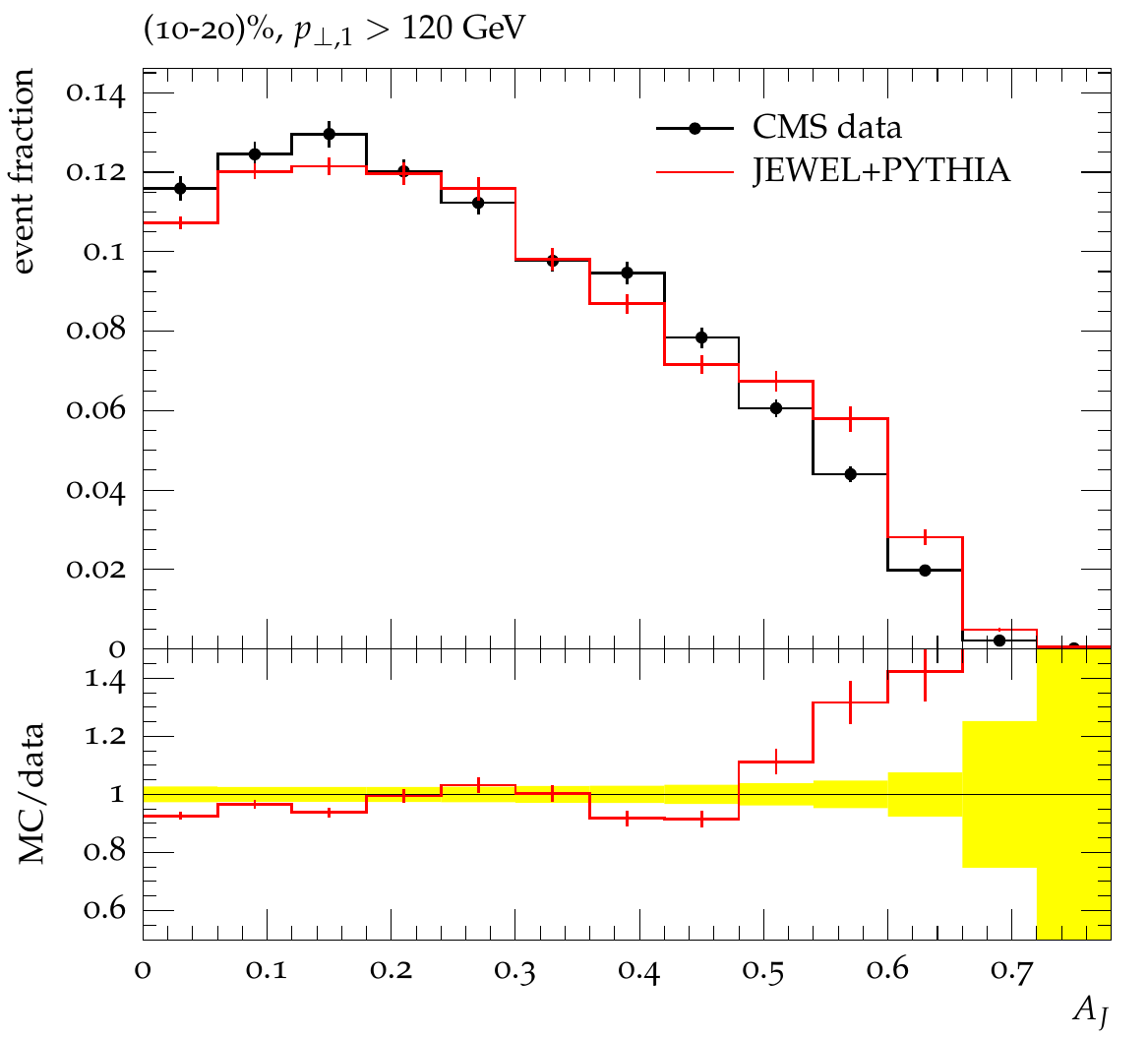}\\
\includegraphics[width=0.45\linewidth]{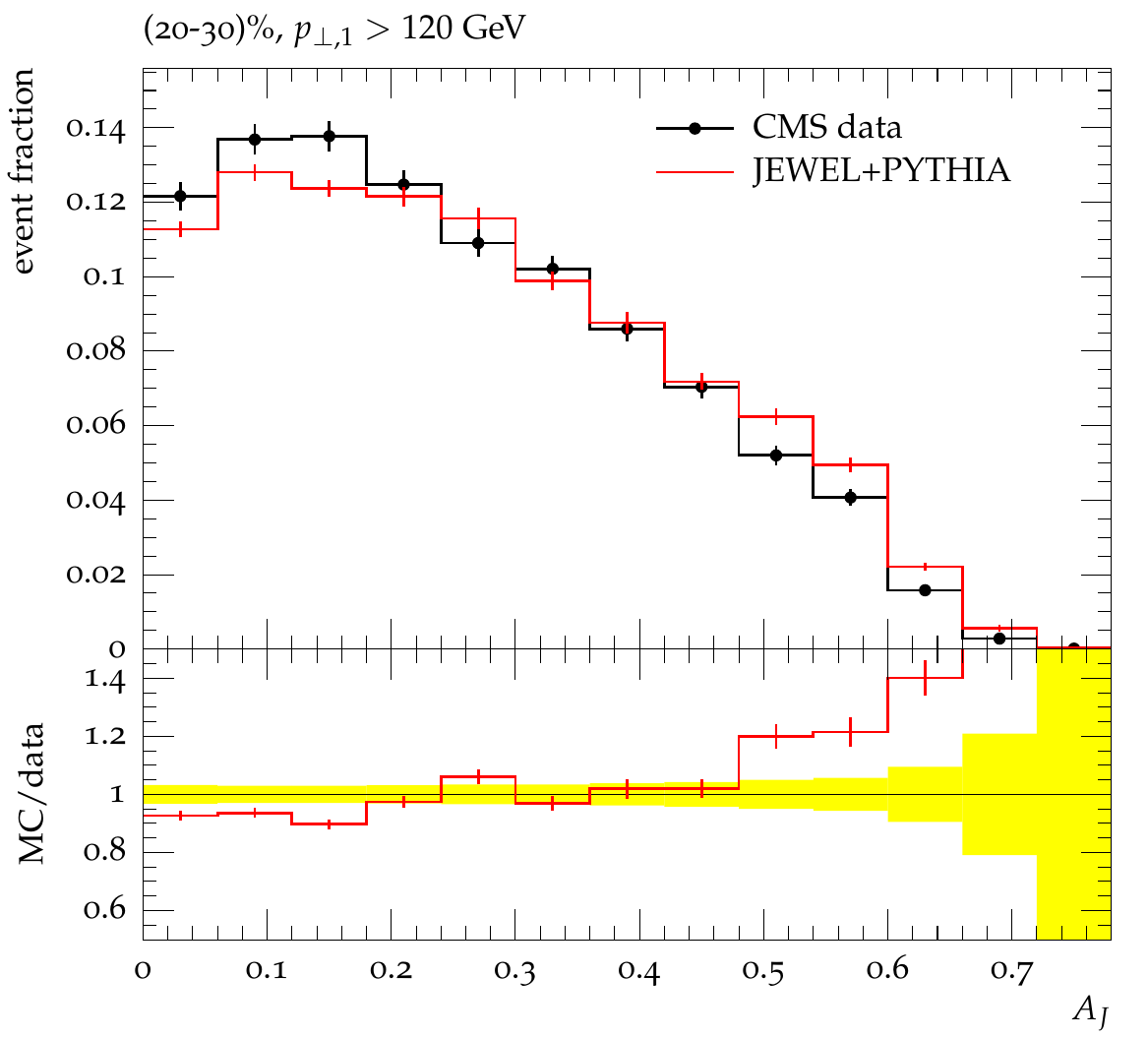}
\includegraphics[width=0.45\linewidth]{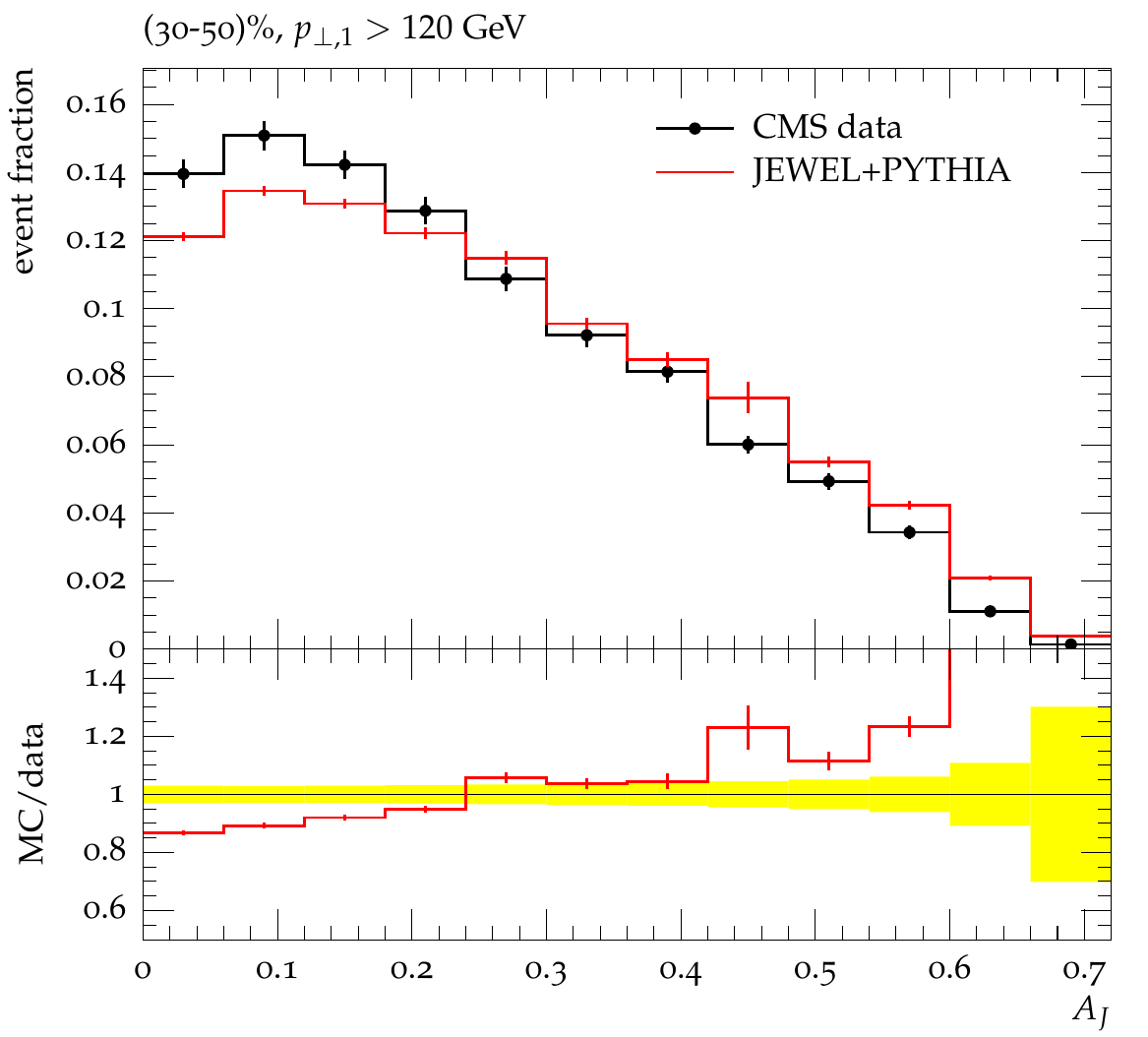}\\
\includegraphics[width=0.45\linewidth]{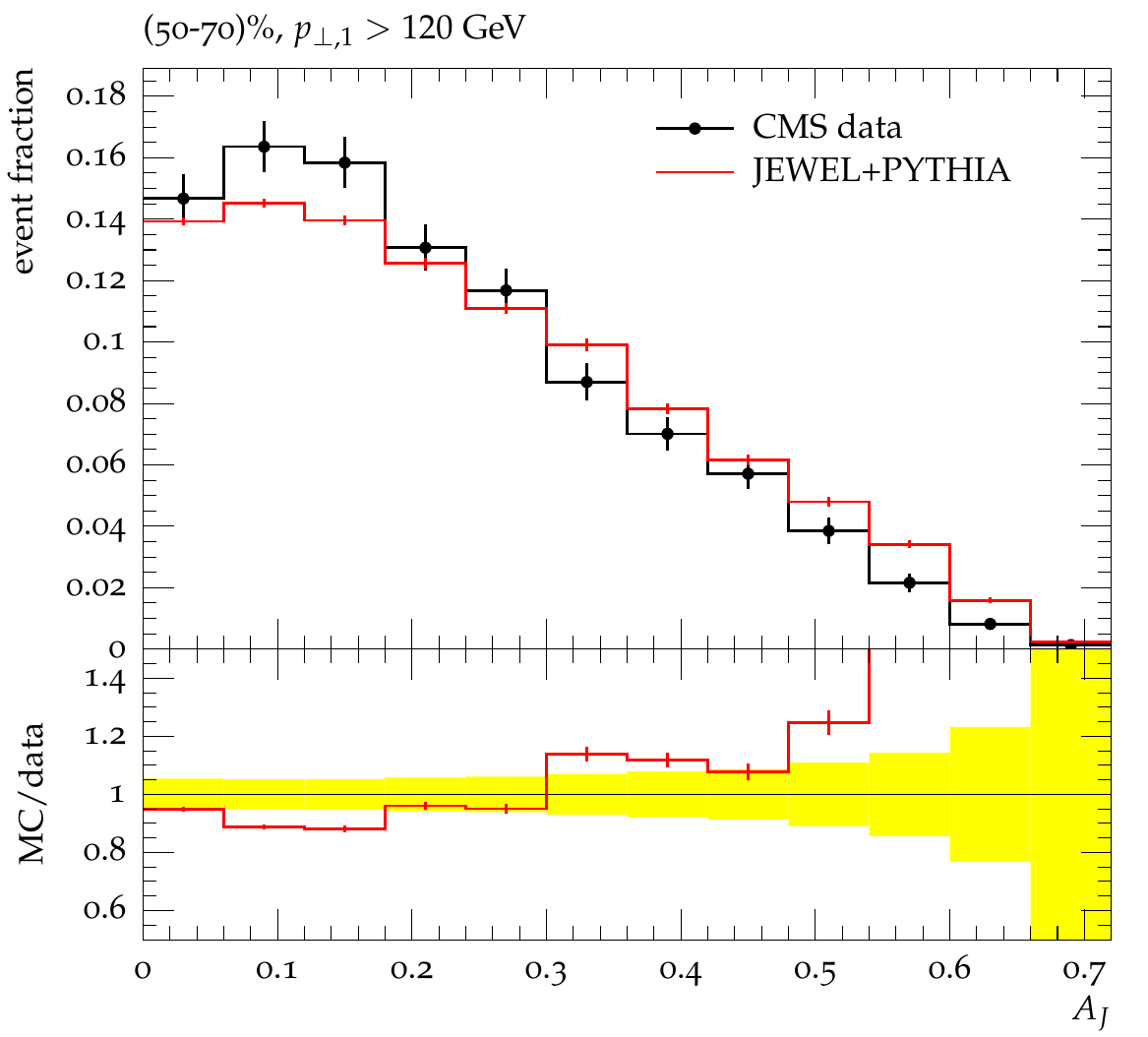}
\caption{Centrality dependence of the di-jet $\pt$-asymmetry $A_\text{J} = (p_{\perp,1} + p_{\perp,2})/(p_{\perp,1} - p_{\perp,12})$ in Pb+Pb collisions  at $\sqrt{s_\text{NN}} = \unit[2.76]{TeV}$ with a jet radius $R=0.3$ and $|\eta_\text{jet}| < 2$~\cite{Chatrchyan:2012nia}. The leading jet has $p_{\perp,1} > \unit[120]{GeV}$ while the sub-leading jet is required to have $p_{\perp,2} > \unit[30]{GeV}$ and $\Delta \phi > 2\pi/3$.  The data are not unfolded for jet energy resolution, so the Monte Carlo events were smeared with the parametrisation from~\cite{Chatrchyan:2012gt}.}
\label{Fig::aj}
\end{figure}

\begin{figure}
\centering
\includegraphics[width=0.45\linewidth]{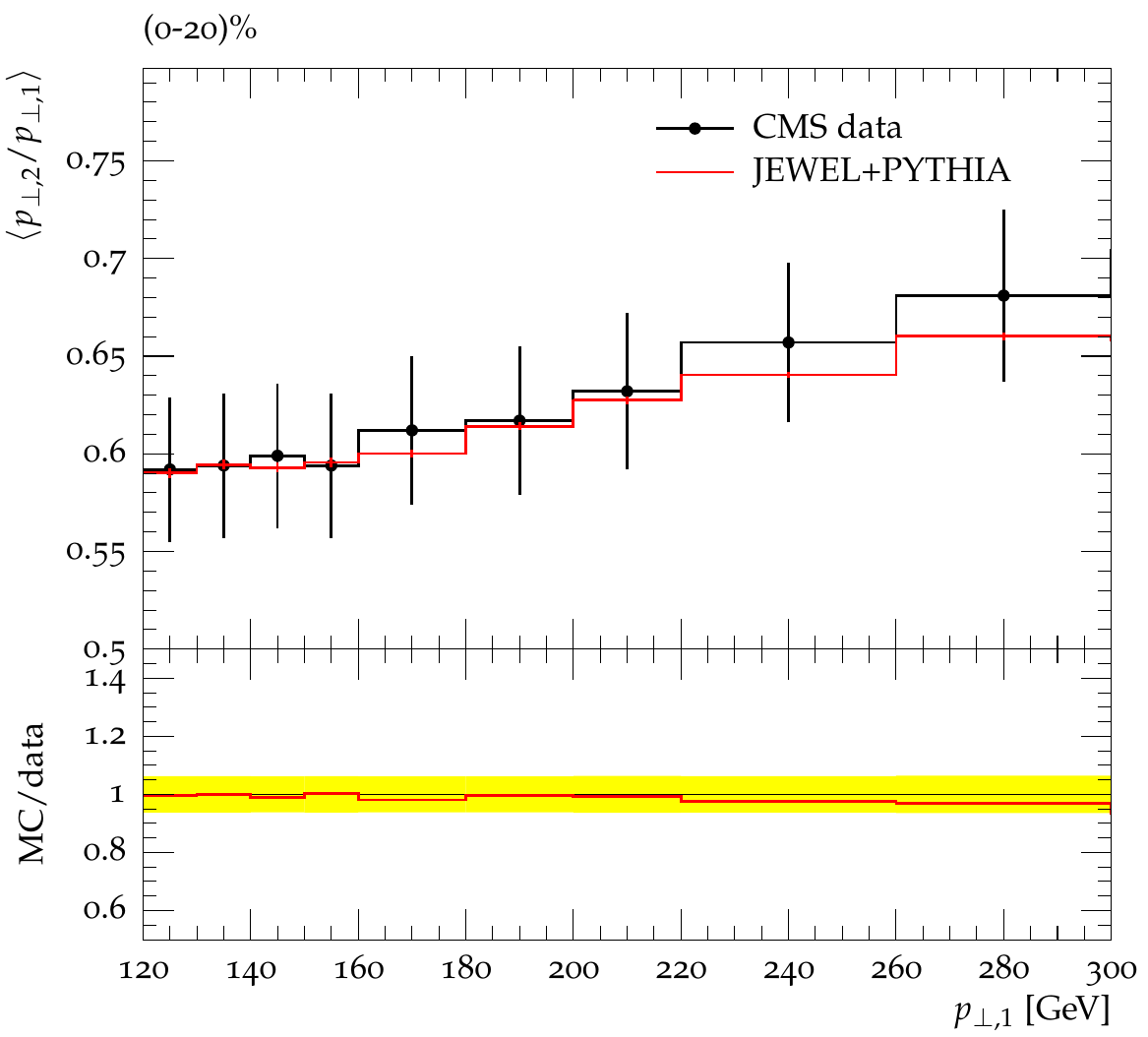}
\includegraphics[width=0.45\linewidth]{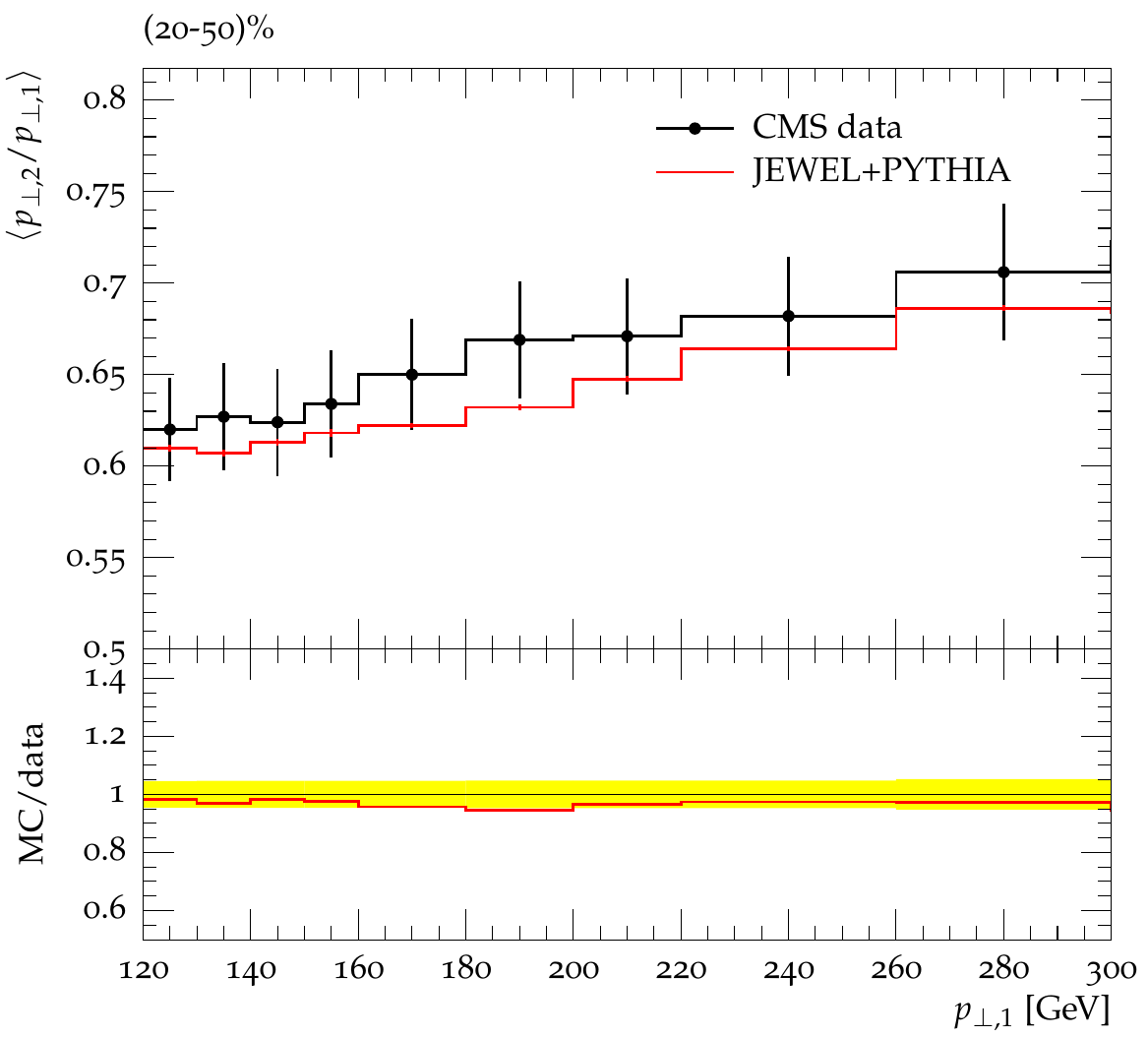}
\caption{Two centrality bins of the mean $\pt$-ratio in di-jet events in  Pb+Pb collisions at $\sqrt{s_\text{NN}} = \unit[2.76]{TeV}$ as a function of the transverse momentum of the leading jet with a jet radius $R=0.3$ and $|\eta_\text{jet}| < 2$~\cite{Chatrchyan:2012nia}. The sub-leading jet is required to have $p_{\perp,2} > \unit[30]{GeV}$ and $\Delta \phi > 2\pi/3$. The data are not unfolded for jet energy resolution, so the Monte Carlo events were smeared with the parametrisation from~\cite{Chatrchyan:2012gt}.}
\label{Fig::ptratio}
\end{figure}

The $\pt$ asymmetry (figure~\ref{Fig::aj}) and the mean $\pt$ ratio (figure~\ref{Fig::ptratio}) in di-jet events are also well reproduced by \textsc{Jewel+Pythia}. The Monte Carlo somewhat overshoots the data in the region of very large asymmetries, but the quality of the agreement does not depend on centrality. The mean $\pt$ ratio is described very well by the Monte Carlo, but here only a few centrality bins have been measured.

\begin{figure}
\centering
\includegraphics[width=0.45\linewidth]{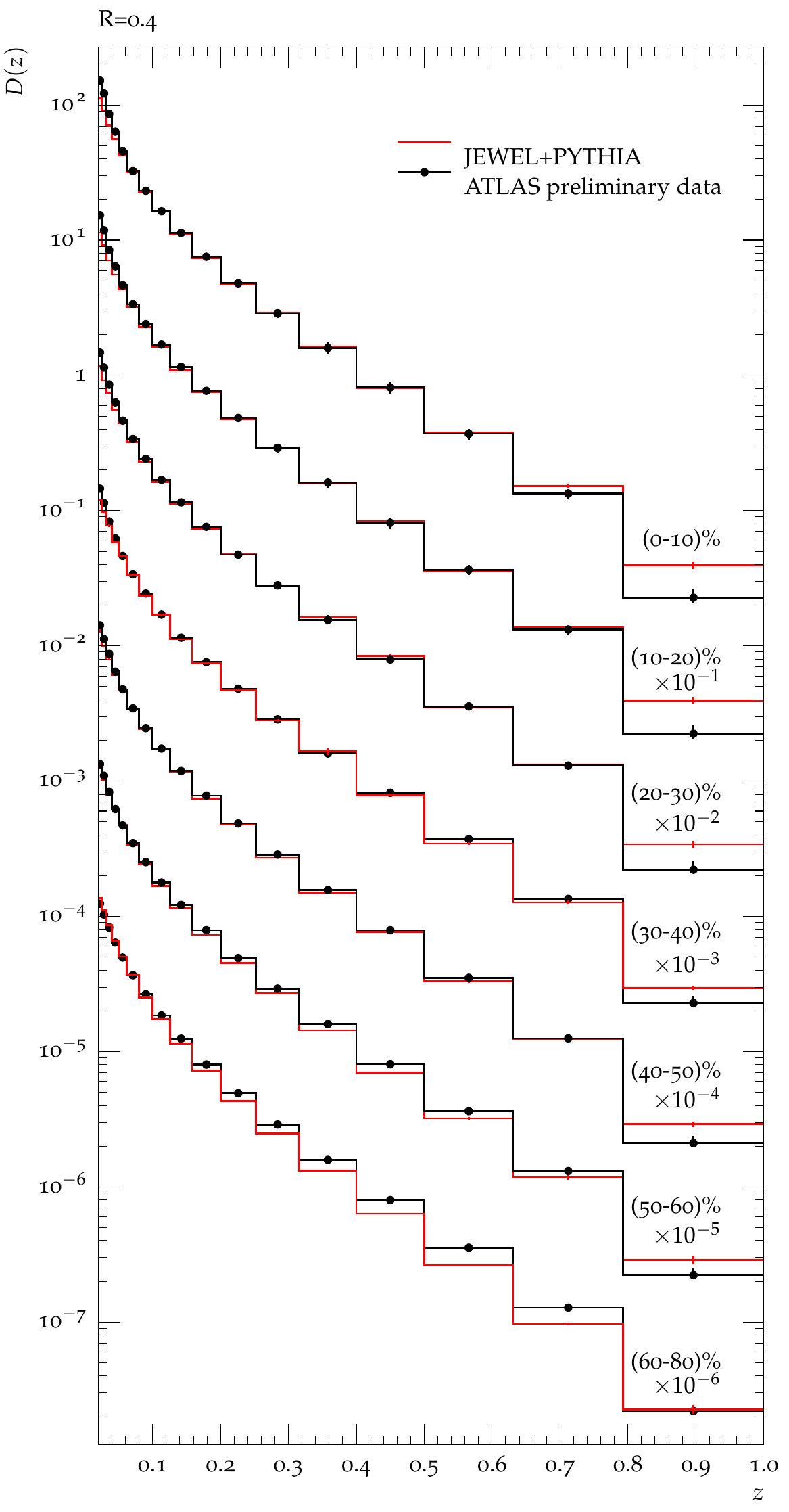}
\includegraphics[width=0.45\linewidth]{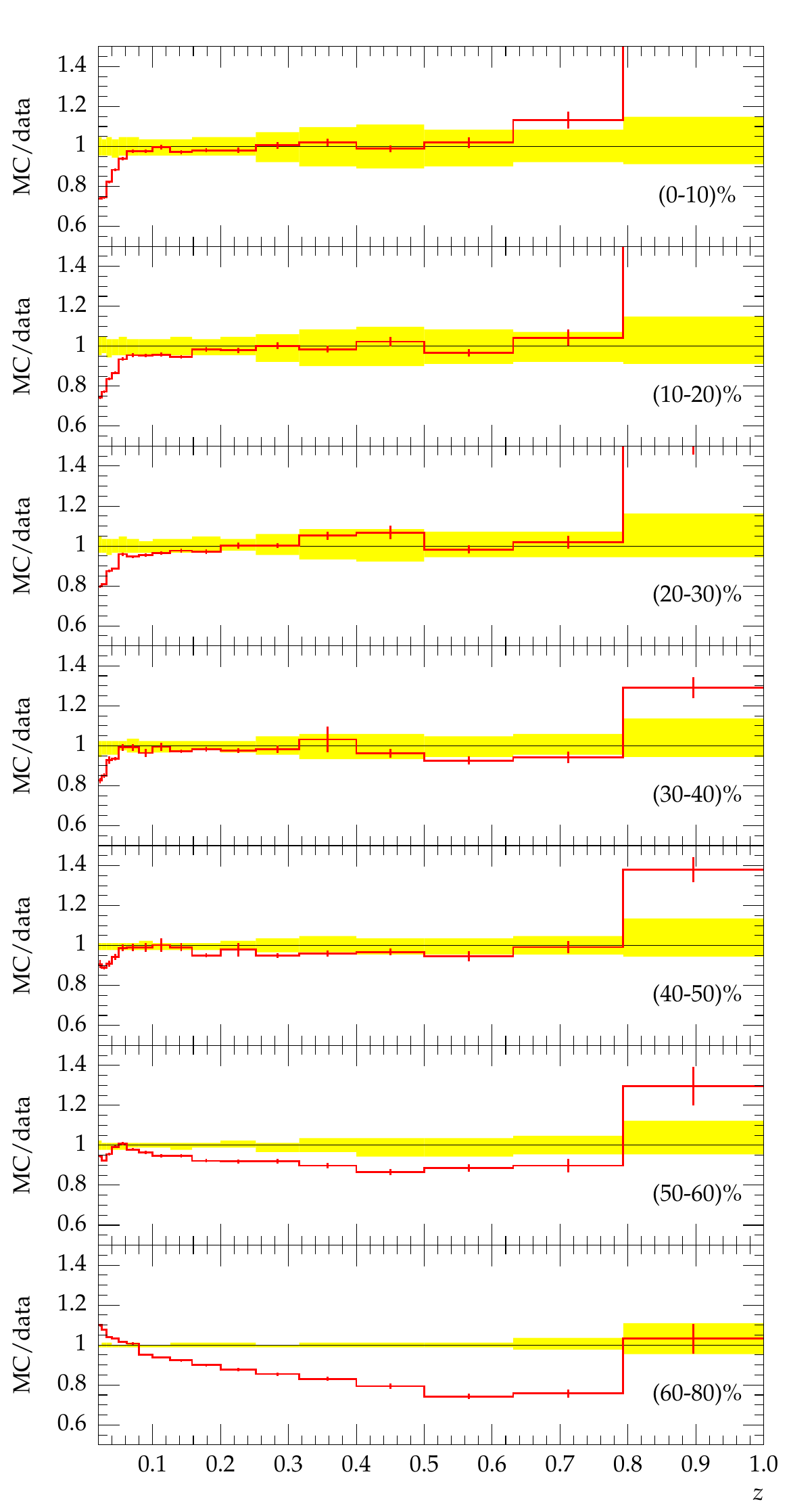}
\caption{Centrality dependence of the intra-jet fragmentation function $D(z)$ in  Pb+Pb collisions at $\sqrt{s_\text{NN}} = \unit[2.76]{TeV}$ for a jet radius $R=0.4$, $p_{\perp,\text{jet}} > \unit[100]{GeV}$ and $|\eta_\text{jet}| < 2.1$~\cite{ATLAS-CONF-2012-115} (data points read off the plots, only maximum of statistical an uncorrelated systematic errors shown).}
\label{Fig::ffs}
\end{figure}

Finally, the intra-jet fragmentation functions $D(z)$ (figure~\ref{Fig::ffs}) also show a reasonable agreement between data and \textsc{Jewel+Pythia}. Only the low $z$ region, which is very susceptible to the treatment of background and therefore differences in the analysis of data and Monte Carlo, and the last bin are not well reproduced. In the last two centrality classes (\unit[50-60]{\%} and \unit[60-80]{\%}) the \textsc{Jewel+Pythia} result starts falling below the ATLAS data, but even in the most peripheral class the largest deviation is about \unit[25]{\%}. This is consistent with the observation in~\cite{Zapp:2012ak} that the \textsc{Jewel+Pythia} fragmentation is too soft in p+p collisions at the LHC. For the central and mid-central bins the Monte Carlo follows the centrality dependence of the data nicely. The agreement is even slightly better for the fragmentation function $D(\pt)$ (not shown).

\section{Conclusions}

In summary, we have extended for the first time studies of jet quenching with \jewel to non-central collisions. This does not introduce any new freedom: the centrality dependence and geometry are encoded in the medium model and the infra-red regulator, which is the only parameter not entirely fixed by other contraints, is adjusted in the most central class of events.

The \jewel framework describes the interactions of jets in a dense and hot medium using standard perturbative technologies and is based on a minimal set of assumptions. It can in principle be interfaced to any model of the medium. However, as this is a novel approach, it has so far been explored with a simple  model of the medium. The initial conditions comprising also the entire centrality dependence are calculated in a simple Glauber model. It follows a hydrodynamic phase of boost-invariant longitudinal expansion with an ideal gas equation of state. It has been shown that with this model \jewel describes a large variety of jet observables in central collisions on a rather satisfactory level~\cite{Zapp:2012ak,Zapp:2013vla}. In this study the performance in non-central collisions, which is a prediction in the sense that there is no freedom to tune parameters, is investigated. Despite the simplistic model of the medium the single-inclusive jet suppression, angular distribution, di-jet asymmetry and intra-jet fragmentation function are described very reasonably over the entire centrality range with no visible systematic trend. The excellent agreement with the jet angular distribution may be somewhat surprising. However, in the \jewel framework different effects caused by transverse expansion, namely faster dilution, restricted phase space and Lorentz contraction, are expected to partially cancel so that this is maybe not accidental. This will be investigated further in an upcoming publication. So far it can only be concluded that \jewel with the simple model of the medium describes a large variety of jet data for central as well an non-central collisions at the LHC reasonably well.

\section*{Acknowledgements}
 The author would like to thank F.~Krauss and U.~Wiedemann for fruitful discussions and comments on the manuscript.

\bibliographystyle{elsarticle-num}
\bibliography{../bib/jetquenching}
\end{document}